\documentclass[a4paper,12pt]{article}
\usepackage{graphicx}
\usepackage{amsmath,amssymb}
\usepackage{indentfirst}

\newcommand{\Per}{{\rm Per}}
\newcommand{\Imm}{{\rm Imm}}
\newcommand{\w}{{\rm W}}

\newtheorem{proposition}{Proposition}
\newtheorem{conjecture}{Conjecture}

\newcommand{\be}{\begin{equation}}
\newcommand{\ee}{\end{equation}}

\newcommand{\U}{\mathcal{U}}
\renewcommand{\O}{\mathcal{O}}

\begin{document}

\title{On the immanants of blocks from random matrices in some unitary ensembles}
\author{Lucas H. Oliveira, Marcel Novaes \\ Instituto de F\'isica, Universidade Federal de Uberl\^andia\\ Uberl\^andia, MG, 38408-100, Brazil}
\date{}

\maketitle
\begin{abstract}
The permanent of unitary matrices and their blocks has attracted increasing attention in quantum physics and quantum computation because of connections with the Hong-Ou-Mandel effect and the Boson Sampling problem. In that context, it would be useful to know the distribution of the permanent and other immanants for random matrices, but that seems a difficult problem. We advance this program by calculating the average of the squared modulus of a generic immanant for blocks from random matrices in the unitary group, in the orthogonal group and in the circular orthogonal ensemble. In the case of the permanent in the unitary group, we also compute the variance. Our approach is based on Weingarten functions and factorizations of permutations. In the course of our calculations we are led to a conjecture relating dimensions of irreducible representations of the orthogonal group to the value of zonal polynomials at the identity.

\end{abstract}

\section{Introduction and Results}

With an $n\times n$ matrix $A$ we can associate some quantities called immanants, which are labelled by partitions of $n$. The most important and most studied of them is by far the determinant. This is because the determinant is invariant under similarity transformations and can be expressed solely in terms of the eigenvalues (this is not true of the other immanants). Given the partition $\gamma\vdash n$, the corresponding immanant is given by
\be 
\Imm_\gamma (A) = \sum_{\pi \in S_n} \chi_\gamma(\pi)\prod^n_{k=1}A_{k,\pi(k)},
\ee 
where the sum is over the permutation group $S_n$ of $n!$ elements, and $\chi_\lambda(\pi)$ is the character of $\pi$ in the irreducible representation of $S_n$ labelled by $\gamma$. The determinant corresponds to the partition with all parts equal to $1$, which leads to the totally antisymmetric or alternating representation, for which $\chi_{(1^n)}(\pi)$ equals the sign of the permutation $\pi$.

The second most famous immanant is the permanent, which seems to have been introduced by Cauchy and is sometimes called the unsigned determinant. It corresponds to the one-part partition $\gamma=(n)$, associated with the totally symmetric or trivial representation, for which $\chi_{(n)}(\pi)=1$. Other immanants, introduced by Littlewood and Richardson in the same paper where they defined their famous rule \cite{LR}, do not have special names.

The permanent has attracted some attention recently within the area of quantum physics, because it is involved in the description of the output state of several bosons scattered by a linear interferometer \cite{Scheel}. This process may be used to implement a (non-universal) platform for quantum computation \cite{aronson} and to solve problems that are intractable using classical computers. In particular, the calculation of the permanent for a generic matrix is supposed to be a \#P-hard problem \cite{valiant} (see also recent experimental results in \cite{exp} and the discussion in \cite{forr}). Permanents of random matrices have been attacked from different directions \cite{perms1,perms2,perms3,fyodorov}.

In the quantum mechanical setting, when the interferometer is modelled by a chaotic cavity, as in \cite{urbina}, it is natural to replace the scattering matrix by a random unitary matrix, uniformly distributed in the unitary group $\U(N)$ with respect to Haar measure. In the presence of time-reversal symmetry, the matrix is additionally taken to be symmetric, i.e. uniformly distributed in the symmetric space $\U(N)/\O(N)$ (also called the Circular Orthogonal Ensemble, COE), where $\O(N)$ is the orthogonal group . 

Motivated by this quantum physics context, in this work we address to problem of computing the average value of immanants of blocks inside random matrices in these three unitary ensembles (for simplicity, we do not discuss symplectic ensembles, but their analysis follows similar lines).
When we write $\Imm_\gamma(U)$ with $\gamma$ a partition of $n$, we understand that we are computing the immanant of the upper-left $n\times n$ block cut out from an $N\times N$ matrix $U$. The determinant and permanent of the block are denoted ${\rm Det}_n$ and $\Per_n$, respectively. We may of course consider the whole matrix by taking $n=N$. 

Some particular results have appeared before. Working in the context of the Hong-Ou-Mandel effect, Urbina {\it et.al.} \cite{urbina} found the average of the squared modulus of the permanent and of the determinant, for the unitary group and for the COE. However, their COE calculation is for blocks whose row indices and columns indices are distinct, while ours is for a diagonal block. Also, Fyodorov \cite{fyodorov}, computed the average of a product of two permanent polynomials of the whole matrix, in the unitary group $\langle \Per_N(U-z_1)\Per_N(U^\dag-z_2)\rangle_{\U(N)}$. We rederive his result by a different method, and generalize it to sub-blocks of size $n$ and to the orthogonal group.

We now briefly review some definitions related to representation theory in order to state our results. Detailed discussion of the concepts involved can be found in Section 2. 

We need some families of monic polynomials, labelled by integer partitions. They are
\be\label{Nl} [N]_\lambda^{(\alpha)}=\prod_{(i,j)\in\lambda}(N+\alpha(j-1)-i+1),\ee 
and
\be\{N\}_\lambda=\prod_{(i,j)\in\lambda}(N-1+e(i,j)),\ee
where $(i,j)$ refers to a box in the Young diagram representation of $\lambda$, and \be e(i,j)=\begin{cases} \lambda_i+\lambda_j-i-j+1,\text{ if }i\le j,\\
-\tilde{\lambda}_i-\tilde{\lambda}_j+i+j-1, \text{ if }i> j \end{cases}\ee (here $\tilde{\lambda}$ is the conjugate partition).
The first family has a parameter $\alpha$. When $\alpha=1$ they are given by 
 \be [N]_\lambda^{(1)}=\frac{n!}{d_\lambda}s_\lambda(1^N),\ee where $d_\lambda=\chi_\lambda(1)$ are dimensions of the irreducible representations of the permutation group and $s_\lambda$ are Schur functions, irreducible characters of $\mathcal{U}(N)$. When $\alpha=2$ they are related to zonal polynomials, \be [N]_\lambda^{(2)}=Z_\lambda(1^N).\ee The other family is also given by \be \{N\}_\lambda=\frac{n!}{d_\lambda}o_\lambda(1^N),\ee where $o_\lambda$ are irreducible characters of $\mathcal{O}(N)$. 

An important role will be played by the function 
\be G_{\lambda,\gamma}= \sum_{\mu\vdash n}|\mathcal{C}_\mu|\omega_\lambda(\mu)\chi_\gamma(\mu),\ee
where $\mathcal{C}_\mu$ are the conjugacy classes of $S_n$ and $\omega_\lambda(\mu)$ are zonal spherical functions of the Gelfand pair $(S_{2n},H_n)$, the group $H_n$ being the hyperoctahedral. The function $ G_{\lambda,\gamma}$ looks like an inner product between a permutation group character and a zonal spherical function, and hence looks somewhat unnatural, mixing up quantities that do not belong together. Nevertheless it appears in our results, in particular in the form of the special case
\be\label{gg} g_\lambda=G_{\lambda,(n)}=\sum_{\mu\vdash n}|\mathcal{C}_\mu|\omega_\lambda(\mu).\ee
We show in Section 2 that the function $g_\lambda$ is different from zero only if $\lambda$ has at most 2 parts and \be g_{(\lambda_1,\lambda_2)}=\frac{(2\lambda_2)!\lambda_1!}{4^{\lambda_2}\lambda_2!}\ee (the calculation of $g_\lambda$ was communicated to us by Sho Matsumoto).

Our first result is 

\begin{proposition} Let $\Imm_\gamma(U)$ denote the $\gamma$-immanant of the $n-$block of the matrix $U$ and $\mathcal{U}(N)$ be the unitary group. Then,
\be\label{P1} \left\langle \left|\Imm_\gamma(U)\right|^2\right\rangle_{\mathcal{U}(N)}=\frac{n!}{[N]_{\gamma}^{(1)}}.\ee
\end{proposition}

This generalizes the special cases of the permanent, $\left\langle \left|\Per_n(U)\right|^2\right\rangle_{\mathcal{U}(N)}=\frac{n!(N-1)!}{(N+n-1)!}$, and of the determinant, $ \left\langle \left|{\rm Det}_n(U)\right|^2\right\rangle_{\mathcal{U}(N)}=\frac{n!(N-n)!}{N!}$. which appeared in \cite{urbina}.

For the unitary group, we also compute the next moment of $\left|\Per_n(U)\right|^2$ in our

\begin{proposition} Let $\Per_n(U)$ denote the permanent of the $n-$block of the matrix $U$ and let $g_\lambda$ be given in (\ref{gg}). Then,
\be\label{p4} \left\langle \left|\Per_n(U)\right|^4\right\rangle_{\mathcal{U}(N)}=\frac{(2^nn!)^2}{(2n)!}\sum_{\lambda\vdash n}\frac{d_{2\lambda}}{[N]_{2\lambda}^{(1)}}g_\lambda^2.\ee
\end{proposition}
 
We turn to unitary symmetric matrices in our 

\begin{proposition} Let $\Imm_\gamma(V)$ denote the $\gamma$-immanant of the $n-$block of the matrix $V$ which belongs to $COE(N)$, the symmetric space $\mathcal{U}(N)/\mathcal{O}(N)$. Then,
\be\label{coe1} \left\langle \left|\Imm_\gamma(V)\right|^2\right\rangle_{COE(N)}=\frac{4^nn!}{(2n)!}\sum_{\lambda\vdash n}\frac{d_{2\lambda}}{[N+1]_{\lambda}^{(2)}}G_{\lambda,\gamma}^2.\ee
\end{proposition}

The expression (\ref{coe1}) looks very similar to the expression (\ref{p4}). If we remember that every element of $COE(N)$ can be written as $V=UU^T$ with $U\in\mathcal{U}(N)$, it will not be surprising that a quadratic result in $COE(N)$ resembles a quartic result in $\mathcal{U}(N)$. 

The determinant corresponds to $\gamma=(1^n)$, in which case we have 
\be G_{\lambda,(1^n)}=\frac{(2n)!}{2^nn!}\frac{\delta_{\lambda,(1^n)}}{d_{2\lambda}}\ee
and 
\begin{equation}\left\langle \left|{\rm Det}_n(V)\right|^2\right\rangle_{COE(N)}= \frac{(n+1)!(N-n+1)!}{(N+1)!}. \end{equation}
When $n=N$ we get of course $1$, as is to be expected.

We also consider the orthogonal group, for which we show

\begin{proposition} Let $\Imm_\gamma(O)$ denote the $\gamma$-immanant of the $n-$block of the matrix $O$ which belongs to $\mathcal{O}(N)$, the orthogonal group. Then,
\be \left\langle \Imm_\gamma(O)^2\right\rangle_{\mathcal{O}(N)}=\frac{n!}{d_\gamma}\frac{2^nn!}{(2n)!}\sum_{\lambda\vdash n}\frac{d_{2\lambda}}{[N]_{\lambda}^{(2)}}G_{\lambda,\gamma}.\ee 
\end{proposition}

In the case of the determinant, we have the same result as for the unitary group, $\left\langle {\rm Det}_n(O)^2\right\rangle_{\mathcal{O}(N)}=\frac{n!(N-n)!}{N!}$.

In the course of our calculations, we have come to consider the validity of a rather surprising identity, a relation between $[N]_{\lambda}^{(2)}$ (associated with zonal polynomials) and $\{N\}_{\gamma}$ (associated with irreducible representations of $\O(N)$). This we formulate as our

\begin{conjecture} The following identity holds:
\be\label{C2} \sum_{\lambda\vdash n}\frac{d_{2\lambda}}{[N]_{\lambda}^{(2)}}G_{\lambda,\gamma}=\frac{(2n)!}{2^nn!}\frac{d_\gamma}{\{N\}_{\gamma}}.\ee
\end{conjecture} 

We have checked this conjecture for all $\gamma\vdash n$ with $n\le 6$. It is not clear to us how to interpret this relation within representation theory (it may be somehow related to the fact that zonal polynomials are orthogonal functions on the symmetric space $\mathcal{U}(N)/\mathcal{O}(N)$).

If Conjecture 1 is true, then it implies
\be \left\langle \Imm_\gamma(O)^2\right\rangle_{\mathcal{O}(N)}=\frac{n!}{\{N\}_{\gamma}},\ee
a result that looks similar to the analogous result for the unitary group, Eq.(\ref{P1}).

Finally, we consider average permanent polynomials, $\Per_n(U-z).$ The average value of this quantity is given by $(-z)^n$ for every random matrix ensemble considered above. For the unitary and orthogonal groups, the quadratic version can be computed from the previous results according to

\begin{proposition} For $G=\U(N)$ or $G=\O(N)$, 
\be \left\langle\Per_n(U-z_1)\Per_n(U^\dag-z_2)\right\rangle_G=\sum_{m=0}^n{n \choose m}(z_1z_2)^{n-m}\left\langle\left|\Per_m(U)\right|^2\right\rangle_G.\ee 
\end{proposition}

For the unitary group, the above expression reduces to $\sum_{m=0}^n(z_1z_2)^{n-m}\frac{n!(N-1)!}{(n-m)!(N+m-1)!},$ which generalizes the $n=N$ result of \cite{fyodorov}.

Results for $\U(N)$, $COE(N)$ and $\O(N)$ are proved in Sections 3, 4 and 5, respectively. Conjecture 1 is discussed in Section 5, along with a symplectic analogue. Permanent polynomials are discussed in Section 6. The main ingredients in our calculations are Weingarten functions and some character theoretic results related to enumerating some factorizations of permutations. These preliminaries are reviewed in Section 2. 

\section{Preliminaries}

\subsection{Permutation groups}

A partition is a weakly decreasing sequence of positive integers, $\lambda=(\lambda_1,\lambda_2,\ldots)$. The number of non-zero parts is its length, $\ell(\lambda)$. By $\lambda\vdash n$ or $|\lambda|=n$ we mean $\sum_{i=1}^{\ell(\lambda)}\lambda_i=n$. So $(3,2,2,1)\vdash 8$. We also write $j^m$ if part $j$ appears $m$ times, so $(3,2^2,1)\equiv (3,2,2,1)$. The rank of a partition is defined as $r(\lambda)=|\lambda|-\ell(\lambda)$.

Let $S_n$ be the group of all permutations acting on the set $\{1,...,n\}$. To a given permutation $\pi\in S_n$ we associate its cycle type, the partition of $n$ whose parts are the lengths of the cycles of $\pi$. Permutation $(1\,2\cdots n)$ has cycle type $(n)$, while the identity permutation has cycle type $(1^n)$. The conjugacy class $\mathcal{C}_\lambda$ contains all permutations with cycle type $\lambda$, and its size is $|\mathcal{C}_\lambda|=\frac{n!}{z_\lambda},$ where \be z_\lambda=\prod_j j^{v_j}v_j!,\ee with $v_j(\lambda)$ being the number of times part $j$ appears in $\lambda$.

Irreducible representations of $S_n$ are also labelled by partitions of $n$, and we denote by $\chi_\lambda(\mu)$ the character of a permutation of cycle type $\mu$ in the representation labelled by $\lambda$. The quantity $d_\lambda=\chi_\lambda(1^n)$, the dimension of the representation, is given by \be\label{dimen} d_\lambda=n!\prod_{i=1}^{\ell(\lambda)}\frac{1}{(\lambda_i-i+\ell)!}\prod_{j=i+1}^{\ell(\lambda)}(\lambda_i-\lambda_j+j-i).\ee 

Characters satisfy two orthogonality relations, 
\be \sum_{\mu\vdash
n}\chi_\mu(\lambda)\chi_\mu(\omega)=z_\lambda\delta_{\lambda,\omega}, \quad
\sum_{\lambda\vdash n}\frac{1}{z_\lambda} \chi_\mu(\lambda) \chi_\omega(\lambda)=
\delta_{\mu,\omega}.\ee The latter is generalized as a sum over permutations as \be\label{cg}
\frac{1}{n!}\sum_{\pi\in S_n}\chi_\mu(\pi) \chi_\omega(\pi\sigma)=
\frac{\chi_\omega(\sigma)}{d_\omega}\delta_{\mu,\omega}.\ee
 
A matching on the set $\{1,...,2n\}$ is a collection of $n$ disjoint subsets with two elements each (`blocks'). For example, we call \be\mathfrak{t}:=\{\{1,2\},\{3,4\},..., \{2n-1,2n\}\}\ee the trivial matching. We will consider the group $S_{2n}$ acting on matchings as follows: if block $\{a,b\}$ belongs to matching $\mathfrak{m}$, then block $\{\pi(a),\pi(b)\}$ belongs to $\pi(\mathfrak{m})$. 

The matching $\mathfrak{m}$ can be represented by the unique permutation $\sigma$ that satisfies he equality $\mathfrak{m}=\sigma (\mathfrak{t})$ and the conditions $\sigma(2k-1)<\sigma(2k)$, $1\le k\le n$, and $\sigma(1)<\sigma(3)<\cdots<\sigma(2n-1)$. The set of all $(2n-1)!!=\frac{(2n)!}{2^nn!}$ such permutations we denote by $\mathcal{M}_n$.

The hyperoctahedral group $H_n\subset S_{2n}$, with $|H_n|=2^nn!$ elements, is the centralizer of $\mathfrak{t}$ in $S_{2n}$, i.e. $H_n=\{h\in S_{2n}, h(\mathfrak{t})=\mathfrak{t}\}$. Elements in the coset $S_{2n}/H_n$ may therefore be represented by the same matching. Also important is the double coset $H_n\backslash S_{2n}/H_n$: permutations $\pi$ and $\sigma$ belong to the same double coset if, and only if, $\pi=h_1\sigma h_2$ for some $h_1,h_2\in H_n$. 

The notion of coset type is important in this context. Given a matching $\mathfrak{m}$, let $\mathcal{G}_\mathfrak{m}$ be a graph with $2n$ vertices labelled from $1$ to $2n$, two vertices
being connected by an edge if they belong to the same block in either $\mathfrak{m}$ or $\mathfrak{t}$. Since each vertex belongs to two edges, all connected components of
$\mathcal{G}_\mathfrak{m}$ are cycles of even length. The coset type of $\mathfrak{m}$ is the partition of $n$ whose parts are half the number of edges in the connected components of $\mathcal{G}_\mathfrak{m}$. We denote by ${\rm ct}(\pi)$ the coset type of $\pi$.

Coset type is clearly invariant under multiplication by hyperoctahedral elements; double cosets are thus labelled by partitions of $n$, so that $\pi$ and $\sigma$ belong to the same double coset if, and only if, they have the same coset type. The double coset associated with the partition $\lambda$ is denoted by $K_\lambda$ an its size is 

\begin{equation}
|K_\lambda| = \frac{4^nn!|\mathcal{C}_\lambda|}{2^{l(\lambda)}}.
\end{equation}

\begin{figure}[t]
\center
\includegraphics[scale=0.9,clip]{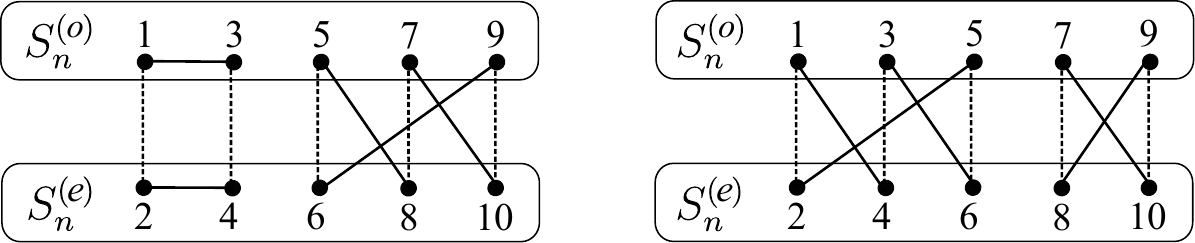}
\caption{The graphs associated with the matchings $\{\{1, 3\}, \{2,4\}, \{5,8\}, \{6,9\}, \{7,10\}\}$ (left) and  $\{\{1, 4\}, \{2,5\}, \{3,6\}, \{7,10\}, \{8,9\}\}$ (right),
whose coset types are both $(3, 2)$. The top and bottom rows are the support of the groups $S_5^{(o)}$ and $S_5^{(e)}$, respectively. The matching on the left can be written as $\sigma(\mathfrak{t})$ with $\sigma=(2\,3)(6\,8\,9\,7)$, while the matching on the right can be written as $\sigma(\mathfrak{t})$ with $\sigma=(2\,4\,5\,3)(8\,10\,9)$. The permutarion $\sigma^\prime = (2\,4\,6)(8\,10) $ acts only on even numbers and also produces the matching on the right.} \label{Fig1}
\end{figure}

In Figure~\ref{Fig1} we show the elements of $\{1,...,2n\}$ arranged in two rows. The top row contains the odd numbers and the bottom row the even ones. The dashed lines indicate how they are paired in the trivial matching. The solid lines come from some other matching $\mathfrak{m}$ and together both sets of lines make up the graph $\mathcal{G}_\mathfrak{m}$.

It will be important to consider two copies of $S_n$ inside $S_{2n}$. The first one acts on odd numbers in the top row of our diagram, while leaving the bottom row invariant, call it $S_n^{(o)}$. The second one acts on even numbers in the bottom row, while leaving the top row invariant, call it $S_n^{(e)}$. For $\pi\in S_n$ we define $\pi_o\in S_n^{(o)}$ and $\pi_e\in S_n^{(e)}$ as
\be \pi_o (2n-1)=2\pi(n)-1, \quad \pi_e (2n)=2\pi(n).\ee
It is easy to see that the coset type of $\pi_o$ and $\pi_e$ are equal to the cycle type of $\pi$:
\be\label{ctct} \pi \in \mathcal{C}_\lambda\Rightarrow {\rm ct}(\pi_o)={\rm ct}(\pi_e)=\lambda.\ee Notice that \be\label{conj} \pi_e=p\pi_o p, \quad p=(1\,2)(3\,4)\cdots (2n-1\,2n).\ee

Another important group is composed of $n$ copies of $S_2$, each generated by transpositions of the kind $(2k-1\,2k)$ with $1\le k\le n$. We denote this commutative group with $2^n$ elements by $S_2^{\otimes n}$. Notice that every element of the hyperoctahedral group has a unique expression as $h=\pi_e\pi_o\xi$, with $\xi \in S_2^{\otimes n}$.

The average \be \omega_\lambda(\tau)=\frac{1}{|H_n|}\sum_{\xi \in
H_n}\chi_{2\lambda}(\tau\xi)\ee is called a zonal spherical function. It is invariant under multiplication by elements of $H_n$ and hence depends only on the coset type of its argument. The simplest cases are $\omega_\lambda(1^{n})=1$ and $\omega_{(n)}(\tau)=1$. They also satisfy orthogonality relations,
\be \label{orthzon}
\sum_{\lambda \vdash n}d_{2\lambda}\omega_\lambda(\alpha)\omega_\lambda(\beta)=\frac{(2n)!}{|K_\alpha|}\delta_{\alpha,\beta},
\ee
and
\be \label{orthzon2}
\sum_{\alpha \vdash n}|K_{\alpha}|\omega_\beta(\alpha)\omega_\lambda(\alpha)=\frac{(2n)!}{d_{2\lambda}}\delta_{\lambda, \beta}.
\ee

\subsection{Factorizations}

An equation of the kind
\be \label{eq}\pi_1\cdots \pi_r=1,\ee
in the permutation group, is called a factorization of the identity. If we specify the cycle types of each factor, $\pi_i \in \mathcal{C}_{\alpha_i}$ then it is a classical result that the number of solutions to the equation is given by
\begin{equation}
\frac{|\mathcal{C}_{\alpha_1}||\mathcal{C}_{\alpha_2}|\cdots |\mathcal{C}_{\alpha_r}|}{n!}\sum_{\beta\vdash n}\frac{\chi_\beta(\alpha_1) \chi_\beta(\alpha_2)\cdots \chi_\beta(\alpha_r)}{d_\beta^{r-2}}.
\end{equation}

When we take the factors as elements of $S_{2n}$ and control their coset types instead of their cycle types, $\pi_i \in K_{\alpha_i}$, there is a similar kind of expression for the number of solutions to equation (\ref{eq}), in which zonal spherical functions take the role previously played by characters \cite{hanlon}: 
\begin{equation}
 F_{\alpha_1\alpha_2\cdots \alpha_r}=\frac{|K_{\alpha_1}||K_{\alpha_2}|\cdots |K_{\alpha_r}|}{(2n)!}\sum_{\beta \vdash n} d_{2\beta}\omega_\beta(\alpha_1) \cdots \omega_\beta(\alpha_r).
\end{equation}

\subsection{Unitary and orthogonal characters}

The celebrated Schur functions $s_\lambda$ are orthogonal characters of the unitary group $\U(N)$,
\be \int_{\U(N)}s_\lambda(U)s_\mu(U^\dag)dU=\delta_{\lambda,\mu}.\ee Their value at the identity matrix is the dimension of the corresponding irreducible representation,
\be s_\lambda(1)=\frac{d_\lambda}{n!}[N]_\lambda^{(1)}.\ee

Zonal polynomials are produced from averages of Schur functions over the orthogonal subgroup,
\be \int_{\mathcal{O}(N)}s_{2\lambda}(AU)dU=\frac{Z_\lambda(A^TA)}{Z_\lambda(1)}.\ee Their value at the identity matrix is known to be
\be Z_\lambda(1)=[N]_\lambda^{(2)}.\ee

The irreducible characters of the orthogonal group are denoted $o_\lambda$, 
\be \int_{\O(N)}o_\lambda(U)o_\mu(U^T)dU=\delta_{\lambda,\mu}.\ee 
Their value at the identity matrix is known to be
\be o_\lambda(1)=\frac{d_\lambda}{n!}\{N\}_\lambda^{(1)}.\ee

\subsection{Weingarten functions}
 
Given $\tau\in S_n$ and two sequences, $\vec{j}=(j_1,\ldots,j_n)$ and $\vec{m}=(m_1,\ldots,m_n)$, define the function \be \delta_\tau[\vec{j},\vec{m}]=\prod_{k=1}^n\delta_{j_{k},m_{\tau(k)}},\ee which is equal to 1 if, and only if, $\vec{m}=\tau(\vec{j})$, i.e. the sequences match up to the permutation $\tau$.

By using these functions, we have \cite{collins,CS}
\begin{equation}
\left\langle U_{a_1b_1} \cdots U_{a_nb_n}U_{c_1d_1}^*\cdots U_{c_nd_n}^* \right\rangle_{\mathcal{U}(N)} = \sum_{\sigma, \tau \in S_n}\delta_\sigma[\vec{a}, \vec{c}]\delta_\tau[\vec{b}, \vec{d}]\w^N_U(\sigma^{-1}\tau),
\end{equation}
where $\w^N_U$ is the Weingarten function of the unitary group, and it is given by
\begin{equation}\label{wu}
\w^N_U(\sigma) = \frac{1}{n!}\sum_{\lambda\vdash  n} \frac{d_\lambda}{[N]^{(1)}_\lambda}\chi_\lambda(\sigma).
\end{equation}

On the other hand, given $\sigma\in \mathcal{M}_{n}$ and $\vec{i} = (i_1,i_2\ldots,i_{2n})$, define the function \be\label{Delta} \Delta_\sigma[\vec{i}]=\prod_{k=1}^n \delta_{i_{\sigma(2k-1)},i_{\sigma(2k)}},\ee which is equal to $1$ if, and only if, the elements of the sequence $\vec{i}$ are pairwise equal according to the matching $\sigma(\mathfrak{t})$. 

For the orthogonal group we have \cite{CS,CM}
\begin{equation}
\left\langle U_{a_1b_1} \cdots U_{a_{2n}b_{2n}} \right\rangle_{\mathcal{O}(N)} = \sum_{\sigma, \tau \in \mathcal{M}_n}\Delta_\sigma[\vec{a}]\Delta_\tau[\vec{b}]\w^N_O(\sigma^{-1}\tau), 
\end{equation}
where $\w^N_O$ is the Weingarten function of the orthogonal group, given by
\begin{equation}\w^N_O(\sigma) = \frac{2^nn!}{(2n)!}\sum_{\lambda\vdash  n} \frac{d_{2\lambda}}{[N]^{(2)}_\lambda}\omega_\lambda(\sigma).
\end{equation}

Moreover, for the circular orthogonal ensemble, we have \cite{matsu1,matsu2}
\begin{equation}
\left\langle  U_{a_1 a_2}\cdots U_{a_{2n-1}a_{2n}}U^*_{b_{1}b_{2}} \cdots U_{b_{2n-1}b_{2n}}^*\right\rangle_{COE(N)} = \sum_{\sigma \in S_{2n}}\delta_\sigma[\vec{a}, \vec{b}]\w^{N+1}_O(\sigma).
\end{equation}

For future reference, let us also define an interleaving operation on sequences:
\be \vec{i}\diamond \vec{j}=(i_1,j_1,i_2,j_2,\ldots,i_n,j_n),\ee
so that, for example,
\be (1,2,\ldots,2n-1,2n)=(1,3,\ldots,2n-1)\diamond(2,4,\ldots,2n).\ee

\subsection{The function $g_\lambda$}

The function $g_\lambda=G_{\lambda,(n)}$ is defined as  
\be\label{g} g_\lambda =\sum_{\mu\vdash n} |\mathcal{C}_{\mu}|\omega_\lambda(\mu).\ee
When $\lambda$ has a single part, it follows from $\omega_{(n)}(\alpha)=1$ that $g_{(n)}=n!$.

In general, its value can be found explicitly if we recall the relation between zonal polynomials and power sum symmetric functions \cite{macdonald},
\be Z_\lambda(x)=2^n\sum_{\mu\vdash n} |\mathcal{C}_{\mu}|\omega_\lambda(\mu)\frac{1}{2^{\ell(\mu)}}p_\mu(x).\ee Taking $x=(1,1)$ we have $p_\mu(x)=2^{\ell(\mu)}$ and $Z_\lambda(1,1)=2^ng_\lambda$. On the other hand, from (\ref{Nl}) we have 
\be Z_\lambda(1,1)=\prod_{(i,j)\in\lambda}(2j-i+1).\ee

If $\lambda$ has more than two parts, the above product vanishes. If $\lambda=(\lambda_1,\lambda_2)$, then we get \be\label{ggg} g_{(\lambda_1,\lambda_2)}=\frac{(2\lambda_2)!\lambda_1!}{4^{\lambda_2}\lambda_2!}.\ee

\section{Unitary group}

\subsection{Proof of Proposition 1}

For simplicity, in this Section we define
\be \mathcal{I}^{\U}_\gamma(N)=\left\langle |\mathrm{Imm}_\gamma(U)|^2 \right\rangle_{\mathcal{U}(N)}.\ee
We begin by expanding,
\begin{align}
\mathcal{I}^{\U}_\gamma & = \left\langle \left\vert \sum_{\pi \in S_n} \prod^n_{i=1}\chi_\gamma(\pi) U_{i\pi(i)} \right\vert^2 \right\rangle_{\mathcal{U}(N)} \\
& = \sum_{\pi_1, \pi_2 \in S_n} \chi_\gamma(\pi_1)\chi_\gamma(\pi_2)\left\langle  \prod^n_{i=1} U_{i\pi_1(i)}U_{i\pi_2(i)}^*    \right\rangle_{\mathcal{U}(N)} \\
\end{align}

In terms of the Weingarten function of the unitary group,
\begin{equation}
\left\langle U_{a_1b_1}...U_{a_nb_n}U_{c_1d_1}^*...U_{c_nd_n}^* \right\rangle_{\mathcal{U}(N)} = \sum_{\sigma, \tau \in S_n}\delta_\sigma[\vec{a}, \vec{c}] \delta_\tau[\vec{b}, \vec{d}]\mathrm{W}_U^N(\sigma^{-1}\tau),
\end{equation}
we have
\begin{align}
\mathcal{I}^{\U}_\gamma(N) & = \sum_{\substack{\pi_1, \pi_2 \in S_n \\ \sigma, \tau \in S_n}} \chi_\gamma(\pi_1)\chi_\gamma(\pi_2)\delta_\sigma[\vec{n},\vec{n}] \delta_\tau[\pi_1(\vec{n}), \pi_2(\vec{n})]\mathrm{W}_U^N(\sigma^{-1}\tau) \\
& = \sum_{\substack{\pi_1, \pi_2 \in S_n \\ \sigma, \tau \in S_n}} \chi_\gamma(\pi_1)\chi_\gamma(\pi_2)\delta_\sigma[\vec{n},\vec{n}] \delta_{\pi_2\tau\pi_1^{-1}}[\vec{n},\vec{n}]\mathrm{W}_U^N(\sigma^{-1}\tau)
\end{align}

The quantities $\delta_\sigma(\vec{n},\vec{n})$ and $\delta_{\pi_2\tau\pi_1^{-1}}(\vec{n},\vec{n})$ are different from zero only if $\sigma$ and $\pi_2\tau\pi_1^{-1}$ are both the identity. Therefore,
\begin{equation}
\mathcal{I}^{\U}_\gamma(N) =\sum_{\pi_1, \pi_2 \in S_n} \chi_\gamma(\pi_1)\chi_\gamma(\pi_2) \mathrm{W}_U^N(\pi_2^{-1}\pi_1).
\end{equation}

Using the character expansion of the Weingarten function and the orthogonality relation in Eq.(\ref{cg}) for the characters, we have
\begin{align}
\mathcal{I}^{\U}_\gamma(N)  & = \frac{1}{n!}\sum_{\lambda \vdash n}\frac{d_\lambda}{[N]_{\lambda}^{(1)}}\sum_{ \pi_2 \in S_n} \chi_\gamma(\pi_2)\left(\sum_{\pi_1 \in S_n}\chi_\gamma(\pi_1) \chi_\lambda(\pi_1\pi_2^{-1})\right) \\
& = \frac{1}{n!}\sum_{\lambda \vdash n}\frac{d_\lambda}{[N]_{\lambda}^{(1)}}\frac{n!\delta_{\lambda,\gamma}}{d_\gamma}\left(\sum_{ \pi_2 \in S_n} \chi_\gamma(\pi_2)\chi_\gamma(\pi_2))\right)\\
& = \frac{n!}{[N]_{\gamma}^{(1)}}.
\end{align}

\subsection{Proof of Proposition 2}

If we restrict attention to the simplest immanant, the permanent, we are able to compute a higher moment. Let $\mathrm{Per}_n(U)$ denote the permanent of the $n\times n$ upper left block of the $N\times N$ unitary matrix $U$. For simplicity of notation, let
\be P_n(N)=\left\langle \left\vert \mathrm{Per}_n(U) \right\vert ^4 \right\rangle_{\mathcal{U}(N)}.\ee
Then,
\be
P_n(N) = \sum_{a, b,c, d \in S_n}\left\langle  \prod^n_{i=1}U_{a(i)i}U_{b(i)i}U_{ic(i)}^*U_{id(i)}^*    \right\rangle_{\mathcal{U}(N)}.
\ee

In terms of the Weingarten function,
\begin{equation}
P_n(N)=\sum_{a,b,c,d\in S_n}\sum_{\sigma,\tau\in S_{2n}} \delta_\sigma[a(\vec{n})\diamond b(\vec{n}),\vec{n}\diamond\vec{n}]
\delta_\tau[\vec{n}\diamond\vec{n},c(\vec{n})\diamond d(\vec{n})]\mathrm{W}_U^N(\sigma^{-1}\tau),
\end{equation}

We can lift the action of $a,b,c,d$ to $S_{2n}$ in terms of the subgroups $S_n^{(o)}$ and $S_n^{(e)}$  according to
\begin{align}
P_n(N) & =\sum_{a,b,c,d\in S_n}\sum_{\sigma,\tau \in S_{2n}} \delta_\sigma[a_ob_e(\vec{n}\diamond\vec{n}),\vec{n}\diamond\vec{n}]\delta_\tau[\vec{n}\diamond\vec{n},c_od_e(\vec{n}\diamond\vec{n})]\mathrm{W}_U^N(\sigma^{-1}\tau)
 \\
& = \sum_{a,b,c,d\in S_n}\sum_{\sigma,\tau \in S_{2n}} \delta_{b_e^{-1}a_o^{-1}\sigma}[\vec{n}\diamond\vec{n},\vec{n}\diamond\vec{n}]
\delta_{\tau c_od_e}[\vec{n}\diamond\vec{n},\vec{n}\diamond\vec{n}]\mathrm{W}_U^N(\sigma^{-1}\tau).
\end{align}

The quantity $\delta_{b_e^{-1}a_o^{-1}\sigma}[\vec{n}\diamond\vec{n},\vec{n}\diamond\vec{n}]$ is different from zero only if the product $b_e^{-1}a_o^{-1}\sigma$ belongs to the group $S_2^{\otimes n}$, as discussed in Section 2. Likewise for $\delta_{\tau c_od_e}[\vec{n}\diamond\vec{n},\vec{n}\diamond\vec{n}]$. Writing
\be b_e^{-1}a_o^{-1}\sigma=\sigma_1^{-1}\in S_2^{\otimes n}, \quad \tau c_od_e=\sigma_2\in S_2^{\otimes n},\ee
we get
\begin{align}
P_n(N) & = \sum_{a,b,c,d\in S_n}\sum_{\sigma_1,\sigma_2\in S^{\otimes n}_2} \mathrm{W}_U^N(\sigma_1b_e^{-1}a_o^{-1}\sigma_2d_e^{-1}c_o^{-1}) \\
& = \sum_{a,b,c,d \in S_n}\sum_{\sigma_1,\sigma_2\in S^{\otimes n}_2} \mathrm{W}_U^N(\sigma_1b_ea_o\sigma_2d_ec_o) ,
\end{align}
where we just replaced some permutations by their inverses.

There always exists some $\rho\in S_n$ such that $a_o=b_o\rho_o$. Likewise we can write $c_o=d_o\theta_o$ for some $\theta\in S_n$. This leads to
\begin{equation}
P_n(N) = \sum_{\rho,b,\theta,d \in S_n}\sum_{\sigma_1,\sigma_2\in S_2^{\otimes n}} \mathrm{W}_U^N(\sigma_1 b_e b_o\rho_o\sigma_2 d_e d_o\theta_o).
\end{equation}

As discussed in Section 2, $\sigma_1b_e b_o=h_1$ and $\sigma_2d_e d_o=h_2 $ belong to the hyperoctahedral group, $H_n$. Then,
\begin{align}
P_n(N) & = \sum_{\rho,\theta\in S_n}\sum_{h_1,h_2\in H_n} \mathrm{W}_U^N(h_1 \rho_o h_2 \theta_o) \\
& = \frac{1}{(2n)!}\sum_{\rho,\theta\in S_n}\sum_{h_1,h_2\in H_n} \sum_{\mu \vdash 2n}d_{\mu}\frac{\chi_\mu(h_1 \rho_o h_2 \theta_o)}{[N]^{(1)}_\mu},
\end{align}
where we have used the explicit form of the unitary Weingarten function as defined in Eq.(\ref{wu}).

We know that $\sum_{h_1 \in H_n} \chi_\mu(h_1\rho_o h_2 \theta_o)$ is different from zero only if $\mu  = 2\lambda$ with $\lambda \vdash n$, in which case it is proportional to the zonal spherical function, by definition. Therefore,
\begin{equation}
\label{soma}
P_n(N) = \frac{2^nn!}{(2n)!}\sum_{\rho,\theta\in S_n}\sum_{h_2\in H_n} \sum_{\lambda \vdash n}d_{2\lambda}\frac{\omega_\lambda(\rho_o h_2 \theta_o)}{[N]^{(1)}_{2\lambda}}.
\end{equation}

The function $\omega_\lambda$ depends only on the coset type of $\eta=\rho_o h_2\theta_o$. Keeping in mind that the coset type of $\rho_o$ equals the cycle type of $\rho$, as in Eq.(\ref{ctct}), and the same holds for $\theta_o$ and $\theta$, we thus ask: how many triples $(\rho_o,\theta_o,h_2)$ are there with $\rho\in \mathcal{C}_{\alpha_1}$, $\theta\in \mathcal{C}_{\alpha_2}$, and $h_2\in H_n=K_{(1^n)}$, such that their product has a given coset type, say $\alpha_3$? This is the same as asking for the number of factorizations of the identity $\eta^{-1} \rho_oh_2\theta_o=1$. The answer to this question is
\be |\mathcal{C}_{\alpha_1}| |\mathcal{C}_{\alpha_2}||K_{\alpha_3}||K_{1^n}| \sum_{\beta
\vdash n} \frac{d_{2\beta}}{(2n)!}\omega_\beta(\alpha_1)\omega_\beta(\alpha_2)\omega_\beta(\alpha_3)\omega_\beta(1^n).\ee

Using that $|K_{1^n}| = 2^n n!$ and $\omega_\beta(1^n)= 1$, we arrive at
\begin{equation}
P_n(N) = \left(\frac{2^nn!}{(2n)!}\right)^2\sum_{\alpha_1, \alpha_2, \beta, \lambda \vdash n} |\mathcal{C}_{\alpha_1}| |\mathcal{C}_{\alpha_2}|\frac{d_{2\beta}d_{2\lambda}}{[N]^{(1)}_{2\lambda}}  \omega_\beta(\alpha_1)\omega_\beta(\alpha_2)\left(\sum_{\alpha_3 \vdash n}|K_{\alpha_3}|\omega_\beta(\alpha_3)\omega_\lambda(\alpha_3)\right).
\end{equation}
Using the orthogonality relation (\ref{orthzon2}), we get
\begin{equation}
P_n(N) = \frac{(2^nn!)^2}{(2n)!}\sum_{\alpha_1, \alpha_2,  \lambda \vdash n} |\mathcal{C}_{\alpha_1}| |\mathcal{C}_{\alpha_2}|\frac{d_{2\lambda}}{[N]^{(1)}_{2\lambda}}  \omega_\lambda(\alpha_1)\omega_\lambda(\alpha_2).
\end{equation}

At this point we use \be g_\lambda =\sum_{\alpha\vdash n} |\mathcal{C}_{\alpha}|\omega_\lambda(\alpha)\ee to finally obtain
\begin{equation}
P_n(N) = \frac{(2^nn!)^2}{(2n)!}\sum_{\lambda \vdash n} \frac{d_{2\lambda}}{[N]^{(1)}_{2\lambda}} g_\lambda^2.
\end{equation}
For example,
\begin{align}
P_1(N) & = \frac {2}{N(N+1)}\\
P_2(N) & = \frac {4(3N^2-N+2)}{N^2(N-1)(N+1)(N+2)(N+3) }  \\
P_3(N) & = \frac {144(N^2+N+4) }{N^2(N-1)(N+1)(N+2)(N+3)(N+4)(N+5) } \\
P_4(N) & = \frac {576( 5N^4+30N^3+127N^2+294N+264)}{N^2(N-1)(N+1)^{2}(N + 2)^{2}(N+3)(N + 4)(N+5)(N+6)(N+7)}
\end{align}

For $N\gg 1$ we can use that $[N]^{(1)}_{2\lambda}\sim N^{2n}$ and the orthogonality relation for zonal spherical functions (\ref{orthzon}), to get
\begin{equation}
P_n(N)\sim \frac{n!}{N^{2n}}\sum_\alpha |\mathcal{C}_\alpha|2^{\ell(\alpha)}=\frac{n!(n+1)!}{N^{2n}}.
\end{equation}
The last equation is obtained by setting $x=2$ in the permutation group cycle index polynomial, \be \sum_{\pi \in S_n}x^{\ell(\pi)}=x(x+1)\cdots(x+n-1).\ee 

\section{Circular orthogonal ensemble}

For simplicity, in this Section we define
\be \mathcal{I}^{C}_\gamma(N)=\left\langle |\mathrm{Imm}_\gamma(U)|^2 \right\rangle_{COE(N)}.\ee
We begin by expanding,
\be
\mathcal{I}^{C}_\gamma(N) = \sum_{a, b\in S_n}\chi_\gamma(a)\chi_\gamma(b)\left\langle   \prod_{i=1}^n U_{a(i)i}U_{i b(i)}^*  \right\rangle_{COE(N)}.
\ee

In terms of Weingarten functions of the circular orthogonal ensemble,
\begin{equation}
\left\langle U_{i_1i_2}...U_{i_{2n-1}i_{2n}}U_{j_1j_2}^*...U_{j_{2n-1}j_{2n}}^* \right\rangle_{COE(N)} = \sum_{\tau \in S_{2n}} \delta_\tau[\vec{i}, \vec{j}] \mathrm{W}^{N+1}_O(\tau).
\end{equation}

This leads to
\begin{equation}
\mathcal{I}^{C}_\gamma(N) = \sum_{a, b\in S_n}\chi_\gamma(a)\chi_\gamma(b)\sum_{\tau \in S_{2n}} \delta_\tau[a(\vec{n})\diamond \vec{n},\vec{n}\diamond b(\vec{n})] \mathrm{W}^{N+1}_O(\tau).
\end{equation}

Again we may lift the action of $a,b$ to $S_{2n}$ in terms of the subgroups $S_n^{(o)}$ and $S_n^{(e)}$. Then the delta function above becomes $\delta_\tau[a_o(\vec{n}\diamond \vec{n}), b_e(\vec{n}\diamond \vec{n}) ] = \delta_{a_o^{-1}\tau b_e}[\vec{n}\diamond \vec{n}, \vec{n}\diamond \vec{n}]$.
This is different from zero only if $a_o^{-1}\tau b_e= \sigma$ belongs to the group $S^{\otimes n}_{2}$, discussed in Section 2. This leads to 
\begin{equation}
\mathcal{I}^{C}_\gamma(N) = \sum_{a,b\in S_n}\sum_{\sigma \in S^{\otimes n}_2}  \chi_\gamma(a)\chi_\gamma(b) \mathrm{W}^{N+1}_O(a_o\sigma b_e^{-1}).
\end{equation}

The Weingarten function depends only on the coset type of $a_o\sigma b_e^{-1}$. So, remembering that the coset type of $a_o$ equals the cycle type of $a$, and the same holds for $b_e$ and $b$, it is necessary to count the number of triples $(a_o,b_e,\sigma)$ such that $a\in \mathcal{C}_{\alpha_1}$, $b \in \mathcal{C}_{\alpha_2}$, $\sigma \in S^{\otimes n}_2$ and $a_o \sigma b_e^{-1} \in K_{\alpha_3}$. This is given by
\be |\mathcal{C}_{\alpha_1}| |\mathcal{C}_{\alpha_2}||K_{\alpha_3}||S^{\otimes n}_2| \sum_{\beta
\vdash n} \frac{d_{2\beta}}{(2n)!}\omega_\beta(\alpha_1)\omega_\beta(\alpha_2)\omega_\beta(\alpha_3)\omega_\beta(1^n).\ee
Using again orthogonality relation of zonal spherical functions, this leads to 
\begin{equation}
\mathcal{I}^{C}_\gamma(N) = \frac{4^nn!}{(2n)!}\sum_{ \lambda\vdash n} \frac{d_{2\lambda}}{[N+1]^{(2)}_\lambda }G_{\lambda, \gamma}^2,
\end{equation}
where \be G_{\lambda, \gamma}= \sum_{\alpha \vdash n} |\mathcal{C}_\alpha| \omega_\lambda(\alpha)\chi_\gamma(\alpha)\ee is a generalization of Eq.(\ref{g}) such that $g_\lambda=G_{\lambda,(n)}$. 

The simplest examples are 
\begin{align}
\mathcal{I}^{C}_{(1)}(N)  & = \frac{2}{ N+1} \\
\mathcal{I}^{C}_{(2)}(N) & = \frac {2(3N + 1)}{ N ( N+1)(N + 3)}\\
\mathcal{I}^{C}_{(3)}(N) & = \frac {24}{ N (N + 3)(N + 5)}\\
\mathcal{I}^{C}_{(4)}(N)  & =\frac {24 (5N^2+20N+23)}{ N( N+1)(N + 2)
(N + 3)( N + 5)( N + 7) }.
\end{align}

Following essentially the same calculation done at the end of Section 3.2, one can show that
\begin{equation}
\mathcal{I}^{C}_{(n)}(N) \sim \frac{(n+1)!}{N^n}+O\left(\frac{1}{N^{n+1}}\right).
\end{equation}

\section{Orthogonal group}

It is obvious that odd moments vanish for any immanant, $\left\langle \left(\mathrm{Imm}_\gamma(U)\right)^{2n+1} \right\rangle_{\mathcal{O}(N)}=0$, so the simplest non-trivial moment is the second. For simplicity, in this Section we define
\be \mathcal{I}^{\O}_\gamma(N)=\left\langle \left(\mathrm{Imm}_\gamma(U)\right)^2 \right\rangle_{\mathcal{O}(N)}.\ee

We start by expanding
\be
\mathcal{I}^{\O}_\gamma(N) = \sum_{a, b\in S_n}\chi_\gamma(a)\chi_\gamma(b)\left\langle   \prod_{i=1}^n U_{i,a(i)}U_{i,b(i)}  \right\rangle_{\mathcal{O}(N)},
\ee
or
\be
\mathcal{I}^{\O}_\gamma(N)= \sum_{a, b\in S_n}\chi_\gamma(a)\chi_\gamma(b)\left\langle   \prod_{i=1}^n U_{a(i),i}U_{a(i),ba(i)}  \right\rangle_{\mathcal{O}(N)}
\ee

Using Weingarten functions of the orthogonal group,
\begin{equation}
\left\langle U_{i_1j_1}...U_{i_{2n}j_{2n}} \right\rangle_{\mathcal{O}(N)} = \sum_{\sigma, \tau \in \mathcal{M}_n} \Delta_\sigma[\vec{i}]\Delta_\tau [\vec{j}] W^{N}_O(\sigma^{-1}\tau),
\end{equation}
will lead to two $\Delta$ functions that are 
\be  
 \Delta_\sigma [a(\vec{n})\diamond a(\vec{n}) ]\ee
and 
\be 
\Delta_\tau[\vec{n}\diamond ba(\vec{n})].\ee
The first one requires that $\sigma$ be the trivial matching. Defining $\pi = ba$, we have
\be
\mathcal{I}^{\O}_\gamma(N)  = \sum_{b, \pi \in S_{n}}\sum_{\tau \in \mathcal{M}_n}\chi_\gamma(b^{-1}\pi)\chi_\gamma(b)\Delta_\tau [\vec{n}\diamond \pi(\vec{n})] W^{N}_O(\tau).\ee
The sum over $b$ can be computed using (\ref{cg}) to give
 \be\mathcal{I}^{\O}_\gamma(N) = \frac{n!}{d_\gamma}\sum_{\pi \in S_{n}}\sum_{\tau \in \mathcal{M}_n}\chi_\gamma(\pi)\Delta_\tau [\vec{n}\diamond \pi(\vec{n})] W^{N}_O(\tau).
\ee

Now, we may introduce $\pi_e$ and write
\be
\mathcal{I}^{\O}_\gamma(N)= \frac{n!}{d_\gamma}\sum_{\pi \in S_{n}}\sum_{\tau \in \mathcal{M}_n}\chi_\gamma(\pi)\Delta_{\pi_e\tau} [\vec{n}\diamond \vec{n}] W^{N}_O(\tau)
\ee
As before, the condition $\Delta_{\pi_e\tau} [\vec{n}\diamond \vec{n}]$ implies that $\pi_e\tau$ must be the trivial matching. For each $\pi$ there is only one $\tau$ that satisfies this condition, and it has the same coset type as $\pi_e$. Hence, 
\begin{equation}
\mathcal{I}^{\O}_\gamma(N) = \frac{n!}{d_\gamma}\sum_{\pi \in S_{n}} \chi_\gamma(\pi) W^{N}_O(\pi_e)
=\frac{n!}{d_\gamma}\sum_{\mu \vdash n} \chi_\gamma(\mu)|\mathcal{C}_\mu| W^{N}_O(\mu).
\end{equation}

From the explicit form of the Weingarten function, we get
\begin{align}
\mathcal{I}^{\O}_\gamma(N) & = \frac{n!}{d_\gamma}\sum_{ \mu \vdash n} |\mathcal{C}_\mu|\chi_\gamma(\mu) \frac{2^nn!}{(2n)!}\sum_{\lambda\vdash n}\frac{d_{2\lambda}}{[N]^{(2)}_\lambda}\omega_\lambda(\mu)\\
& = \frac{n!}{d_\gamma}\frac{2^nn!}{(2n)!} \sum_{\lambda \vdash n} \frac{d_{2\lambda}}{[N]^{(2)}_\lambda}\left(  \sum_{\mu\vdash n}|\mathcal{C}_\mu |\chi_\gamma(\mu) \omega_\lambda(\mu) \right) \\
& = \frac{n!}{d_\gamma}\frac{2^n n!}{(2n)!} \sum_{\lambda \vdash n} \frac{d_{2\lambda}}{[N]^{(2)}_\lambda} G_{\lambda,\gamma}.
\end{align}

Simplest examples for the permanent are
\begin{align}
\mathcal{I}^{\O}_{(1)}(N) & = \frac{1}{N} \\
\mathcal{I}^{\O}_{(2)}(N) & = {\frac {2}{ \left( N-1 \right)  \left( N+2\right) }} \\
\mathcal{I}^{\O}_{(3)}(N) & = {\frac {6
}{N \left( N-1 \right) \left( N + 4 \right)}} \\
\mathcal{I}^{\O}_{(4)}(N) & = {\frac {24}{N    \left( N-1 \right)  \left( N+1 \right) \left( N + 6 \right) }} \\
\end{align}
Using that $\sum_{\lambda\vdash n}d_{2\lambda}\omega_\lambda(\alpha)=\frac{(2n)!}{2^nn!}\delta_{\alpha,1^n}$ we have \be\mathcal{I}^{\O}_{\gamma}(N)\sim \frac{n!}{N^n}+O\left(\frac{1}{N^{n+1}}\right).\ee

Extensive simulations have convinced us that the following identity is true:
\be \sum_{\lambda\vdash n}\frac{d_{2\lambda}}{Z_\lambda(1^N)}G_{\lambda,\gamma}=\frac{(2n)!}{2^nn!}\frac{d_\gamma}{\{N\}_{\gamma}},\ee where $Z_\lambda(1^N)=[N]_{\lambda}^{(2)}$ are zonal polynomials. We put this result forth as a conjecture. If indeed true, it implies that 
\be \mathcal{I}^{\O}_\gamma(N) =\frac{n!}{\{N\}_\gamma}.\ee

Although we do not discuss symplectic ensembles in this work, we have a similar conjecture in that case. Let 
\be G'_{\lambda,\gamma}=\sum_{\mu\vdash n} |\mathcal{C}_\mu|\psi_\lambda(\mu)\chi_\gamma(\mu),\ee
where $\psi_\lambda(\mu)$ are twisted spherical functions of the Gelfand pair $(S_{2n},H_n)$ (see \cite{matsu2}), denote by $\langle N\rangle_{\gamma}=\frac{n!}{d_\gamma}sp_\lambda(1^N)$ some polynomials in $N$ that are proportional to dimensions of irreducible representations of $Sp(2N)$ and let $Z'_\lambda(1^N)=2^n[N]_{\lambda}^{(1/2)}$ be symplectic zonal polynomials. Then the conjecture reads
\be \sum_{\lambda\vdash n}\frac{d_{\lambda\cup\lambda}}{Z'_\lambda(1^N)}G'_{\lambda,\gamma}=\frac{(2n)!}{2^nn!}\frac{d_\gamma}{\langle N\rangle_{\gamma}}.\ee 

\section{Permanent polynomials}

From the definition
\be \Per_n(X)=\sum_{\pi \in S_n} \prod_{i=1}^n X_{i,\pi(i)}\ee it is easy to see that $\Per_n(U-z)$ can be written as
\be \Per_n(U-z)=\sum_{P\subset \{1,\ldots,n\}} \sum_{\pi \in S_P} \prod_{i\in P} U_{i,\pi(i)}(-z)^{n-|P|},\ee
where $P$ is summed over all subsets of $\{1,\ldots,n\}$, the group $S_P$ contains all bijections of $P$ into itself and $|P|$ is the cardinality of $P$.

Using this we have
\be \langle \Per_n(U-z_1)\Per_n(U^\dag-z_2)\rangle_G=\sum_{P_1,P_2\subset \{1,\ldots,n\}}  (-z_1)^{n-|P_1|}(-z_2)^{n-|P_2|}\sum_{\substack{\pi_1 \in S_{P_1}\\ \pi_2 \in S_{P_2}}} E_G(\pi_1,\pi_2)\ee where \be E_G(\pi_1,\pi_2)= \left\langle \prod_{i\in P_1}\prod_{j\in P_2} U_{i,\pi_1(i)}U^*_{j,\pi_2(j)}\right\rangle_G.\ee

Taking into account that the lists $\vec{i}$ and $\vec{j}$ do not any repeated indices, we see that for both the unitary and orthogonal groups the above average is different from zero only if $P_1=P_2=P$. In that case we have \be \sum_{\pi \in S_P}\prod_{i\in P} U_{i,\pi(i)}=\Per_P(U),\ee the permanent of the block from $U$ containing the indices in $P$. But the average value does not depend on the particular indices, so $\langle |\Per_P(U)|^2\rangle_{G}=\langle|\Per_m(U)|^2\rangle_{G}$ if $|P|=m$. Since there are ${n\choose m}$ subsets of size $m$, this finishes the proof for all cases.

\vspace{1cm}
{\bf Acknowledgments}
This project grew out of a conversation with Juan Diego Urbina. We thank Sho Matsumoto for providing the proof of Eq. (\ref{ggg}). M. Novaes was supported by grants 306765/2018-7 and 400906/2016-3 from Conselho Nacional de Pesquisa. L.H. Oliveira was supported by a PhD fellowship from Coordenação de Aperfeiçoamento de Pessoal de N\'ivel Superior.

\end{document}